\documentstyle[12pt,epsfig]{article}
 \newcommand{\p}{{\bf p}}
 \newcommand{\bk}{{\bf k}}
 \newcommand{\bK}{{\bf K}}
 \newcommand{\0}{{\bf 0}}
 \newcommand{\Q}{{\bf Q}}
 \newcommand{\lam}{{\lambda}}
 \newcommand{\ra}{{\rangle}}
 \newcommand{\la}{{\langle}}
 \newcommand{\be}{\begin{equation}}
 \newcommand{\ben}{\begin{eqnarray}}
 \newcommand{\een}{\end{eqnarray}}
 \newcommand{\ee}{\end{equation}}
 \newcommand{\al}{\alpha}
 \newcommand{\po}{\stackrel{o}{\p}}

 \newcommand{\ponot}{\stackrel{\not\,{o}}{\p}}
 \newcommand{\poc}{\stackrel{o}{ p}}
 \newcommand{\ponotc}{\stackrel{\not\,{o}}{p}}
 \newcommand{\plo}{\stackrel{\not\,{o}}{\p\smash{\lefteqn{_{12}}}}\phantom{_1}}
 
 \newcommand{\pmo}{\stackrel{\not\,{o}}{\p\smash{\lefteqn{_1}}}\phantom{1}}
 \newcommand{\pno}{\stackrel{\not\,{o}}{\p\smash{\lefteqn{_2}}}\phantom{1}}
 
  \newcommand{\Ko}{\stackrel{o}{\bK}}

 \newcommand{\Dppo}{\stackrel{+}{D\smash{\lefteqn{^{(1/2)}_{\lam'_{s_1}\lam_{s_1}}}}}\phantom{a1aa}}
 \newcommand{\Dppd}{\stackrel{+}{D\smash{\lefteqn{^{(1/2)}_{\lam'_{s_2}\lam_{s_2}}}}}\phantom{a1aa}}
 
 \newcommand{\DDo}{D\smash{\lefteqn{^{(1/2)}_{\lam_{s_1}\lam'_{s_1}}}}\phantom{aaaa}}
 \newcommand{\DDd}{D\smash{\lefteqn{^{(1/2)}_{\lam_{s_2}\lam'_{s_2}}}}\phantom{aaaa}}

\begin{document}

\begin{center} 
{\bf Three-Quark Wave Function of Nucleon in the 
    Quasipotential Approach }
\\
 T.P.Ilichova, S.G.Shulga
\\[6pt] {\it Francisk
Skaryna Gomel State University, LPP JINR, Dubna}  \\
e-mail: shulga@cv.jinr.ru, ilicheva@cv.jinr.ru
\end{center}

\begin{abstract}
The structure of the nucleon wave function as a bound system
of the constituent quarks was considered in framework of the
quasipotential method of description of the bound states
with a fixed number of particles. 
In the impulse approximation
the wave function is reduced to the three-quark component
of the vector state in the Fock-momentum space.
Spin structure of the wave function was studied by the decomposition method
of the irreducible representations product
of inhomogeneous Lorentz group. Physical distinction of
SU(6) symmetric solution is determined by uniqueness of
this solution in the nonrelativistic limit.
The effective mass
approximation for the relativistic theory was numerically studied,
in which dispersion of the average quark momentum is small
as compare to a big average quark momentum.
Relativistic generalization of the nonrelativistic
three-particle oscillator was proposed.
\end{abstract}

\begin{center}
{\bf Introduction.}
\end{center}

 \par
The two approaches have a leading position among others
to solve a question about the construction
of the relativistic many particle wave function (WF).
The first approach based on many-time 4-dimensional formalizm and
its various
3-dimentional reductions (quasipotential (QP) formulations),
in which WF is defined on equal-time surface in the
center-of-mass system or in laboratory system.
The second one is formulated in light front formalism~\cite{Kondratyuk80}.
Vacuum fluctuations are absent in the light front formalism
and this allows to apply the fixed number particle approximation.
The light front formalism is more suitable for description
at high energies~\cite{Karmanov98}.
 \par
It may be expected that the first approach listed above is preferable
at low and medium energies.
The aim of this work is to demonstrate this property
by using example of description of nucleon in the
quark model in the Logunov-Tavkhelidze
QP approach~\cite{Log-Tav63}, 
in covariant formulation~\cite{Faustov1}.
The vacuum fluctuations, which
are called the main defect of this 
method~\cite{Berest_Terent76},
may be suppressed at low energies.
One task for developing approach consists of the answer 
to the question:
in what region mechanizm of constituent quark may be applied
for the nucleon, 
and that must be solved by using of the experimental data.
\par
The fixed number particles assumption may be 
formally expressed as a zero-order interaction
in quark operators in the Bethe-Salpeter wave function.
This allows one to reduce QP WF
to the decomposition component of the vector state 
in the Fock-momentum space.
Then it is possible to apply an apparatus of decomposition 
of the product of irreducible
representations of the Lorentz-group~\cite{Shirokov58,macfarlane},
which are analogous of the Clebsh-Gordan decomposition 
in quantum mechanics.
Thus the QP method is the analogy of quantum mechanics description, 
which was fruitfully used by many authors. 
  In the third section of
the article this property of the QP approach is used to 
obtain a solution of
the three particle relativistic oscillator.
\newpage
\begin{center}
{\bf 1. Three quark wave function of nucleon
in quasipotential approach.}
\end{center}
\par
Let us suppose that the Fock-momentum space for nucleon 
has a 3-quark basis
and the nucleon vector state in this space decomposes as:
\be
\mid K J \lam_J T \lam_T\ra=
\frac{1}{(2\pi)^9}
\int(\prod\limits_{k=1}^{3}d\Omega_{\p_k})
\mid \{\p_k s_k \lam_{s_k} t_k \lam_{t_k}\}\ra
\la
\{\p_k s_k \lam_{s_k} t_k \lam_{t_k}\}
\mid K J \lam_J T \lam_T\ra.
\label{decomp_1}
\ee
Orthonormalization conditions for the nucleon and quark
vector states are following:
$$
\la K' J' \lam'_J T' \lam'_T
\mid K J \lam_J T \lam_T\ra
=
(2\pi)^3
\delta^3(\bK'-\bK)
\delta_{\lam'_J \lam_J}
\delta_{\lam'_T \lam_T},
$$
$$
\la
\p'_k s_k \lam'_{s_k} t_k \lam'_{t_k}
\mid
\p_k s_k \lam_{s_k} t_k \lam_{t_k}
\ra
=
(2\pi)^3
E_{\p_k}
\delta^3(\p'_k-\p_k)
\delta_{\lam'_{s_k} \lam_{s_k}}
\delta_{\lam'_{t_k} \lam_{t_k}},
$$
\be
\bar{u}_{\al_k}^{\lam_{s_k}}(\p)
u_{\al_k}^{\lam_{s_k}}(\p)
=
m
\quad
(\mbox{there is no sum over }\lam_{s_k}).
\label{normir}
\ee
Here and further $\{\p_k s_k \lam_{s_k} t_k \lam_{t_k}\}$ 
means a set of values with indices k=1,2,3;
$d\Omega_{\p_k}=d\p_k/E_{\p_k}$;
$\p_k,\;m$ are momentum and mass of quark;
$K=(\sqrt{\bK^2+M^2},\bK)$,
$M$ is the nucleon mass;
$J,\;s_k$ are spins and $T,\;t_k$ are isospins of nucleon
and constinuent quarks;
$\lam_x$ is the third projection of x.
In this section the isospin indices will be omitted.
We assume the sum over the repeated spin and isospin indices
(if we don't indicate espessially).
\par
In the quasipotential approach let us consider a
covariant projection
of the Bethe-Salpeter wave function on equal-times surface
in the center-of-mass system~\cite{Log-Tav63,Faustov1}:
\be
\Psi_{KJ\lam_J}^{\{\al_k\}}(\{x_k\})
=
\Psi_{KJ\lam_J}^{BS\{\al_k\}}(\{x_k\})
\delta(n_K(x_1-x_2))
\delta(n_K(x_1-x_3)),
\label{equal_times_WF}
\ee
where
$
\Psi_{KJ\lam_J}^{BS\{\al_k\}}(\{x_k\})=
\la 0\mid T
\left[
\phi_{\al_1}^{(1)}(x_1)
\phi_{\al_2}^{(2)}(x_2)
\phi_{\al_3}^{(3)}(x_3)
\right]
\mid KJ\lam_J\ra
$
is the Bethe-Salpeter wave function, $(n_K)_\mu=K_\mu/\sqrt{K^2}$
is the unit 4-momentum.
$\delta$-functions make the times to be equal 
covariantly in the c.m.s.
Using a property of translational invariance,
it is possible to separate
motion of center-of-mass system 
and then the Fourier-representation
of the wave function~(\ref{equal_times_WF}) 
is given by~\cite{Faustov1},
notation
$\sum_{k=1}^{3}\poc_{k0}=\stackrel{o}{P}_{0}$ is introduced:
\be
\Psi_{KJ\lam_J}^{\{\al_k\}}(\{p_k\})=
\left[
\prod_{k=1}^{3}
S_{\al_k\beta_k}(L_K)
\right]
(2\pi)^4
\delta^{(1)}(\stackrel{o}{P}_{0}-M)
\delta^{(3)}(\sum_{k=1}^3\po_k)
\Phi^{\beta_k}_{MJ\lam_J}(\po_k),
\label{fourie_equal_times_WF}
\ee
where
$
S_{\al_k\beta_k}(L_K)
$ is a matrix of the Lorentz-boost representation  $L_K$,
$L_K^{-1}K\equiv\stackrel{o}{K}=(M,\bf{0})$,
$\poc_k=L^{-1}_K p_k$. 
In~(\ref{fourie_equal_times_WF})
the wave function depends on two independent 3-momenta.
For the sake of symmetry we keep all
the momenta assuming their relation $\sum_{k=1}^3\po_k=0$.
The expression for $\Phi$ has the form:
\be
\Phi^{\beta_k}_{MJ\lam_J}(\po_k)=
\int
[\prod_{k=2}^3d\stackrel{o}{y}_k
\exp(-i\po_k\stackrel{o}{y}_k)]
\la 0\mid
\phi_{\beta_1}^{(1)}(0)
\phi_{\beta_2}^{(2)}(0,\stackrel{o}{y}_2)
\phi_{\beta_3}^{(3)}(0,\stackrel{o}{y}_3)
\mid M \stackrel{o}{\bK}J\lam_J\ra.
\label{phi}
\ee
In the quantum-field interpretation the fixed number particles 
assumption is expressed in replacement of the Heizenberg 
field operators to the Dirac ones
(zero-order interaction in field operators). 
Bound effects are contained
in the Heizenberg vector state of the bound system. 
We have applied this assumption
in~(\ref{phi}) and substituted it
in~(\ref{fourie_equal_times_WF})
and as a result the wave function~(\ref{fourie_equal_times_WF})
is expressed through the decomposition component~(\ref{decomp_1})
in the form:
\be
\Psi_{KJ\lam_J}^{\{\al_k\}}(\{p_k\})=
(2\pi)
\delta^{(1)}(\stackrel{o}{P}_{0}-M)
\left[
\prod_{k=1}^{3}
u^{(\lam'_{s_k})}_{\al_k}(\p_k)
\stackrel{+}{D}\smash{^{1/2}_{\lam'_{s_k}\lam_{s_k}}}(L^{-1}_K,p_k)
\right]
\la
\{\po_k s_k \lam_{s_k}\}
\mid M \stackrel{o}{\bK} J \lam_J \ra.
\label{psi_chrz_compon}
\ee
For obtaining~(\ref{psi_chrz_compon}) we have used the relations:
$$
S_{r i}(L_K)u^{(\lam_{s_k})}_i(\po_k)
=u^{(\lam'_{s_k})}_{r}(\p_k)
\stackrel{+}{D}\smash{^{1/2}_{\lam'_{s_k}\lam_{s_k}}}(L^{-1}_K,p_k),
$$
\be
\hat{U}(L^{-1}_K)
|\p_k s_k \lam_{s_k}\ra
=
|\po_k s_k \lam'_{s_k}\ra
D\smash{^{1/2}_{\lam'_{s_k}\lam_{s_k}}}(L^{-1}_K,p_k),
\label{vig_pov1}
\ee
where $
D^{1/2}(L^{-1}_K, p_k)$
is the Wigner spin rotation matrix,
which is defined as a $2\times2$
representation of the 3-dimensional rotation
$R^W_{(L^{-1}_K, p)}=L^{-1}_{\po} L^{-1}_K L_p$
~\cite{Shirokov58,macfarlane,Shirokov54,Gaziorovich}.
It has an explicit form~\cite{Shirokov58}:
$$
D(L^{-1}_K, p_k)=
\frac
{(E_{\p_k}+m)(E^M_K+M)-(\sigma\bK)(\sigma\p_k)}
{\sqrt{
2(E_{\p_k}+m)(E^M_K+M)(E_{\bK} E_{\p_k}-\bK\p_k+mM)
}}$$

We multiply~(\ref{psi_chrz_compon}) by
$
\prod_{k=1}^{3}
(1/m)
\bar{u}^{(\lam_{s_k})}_{\al_k}(\p_k)
$ from left-handed side and sum the result over the
spinor index:
\be
\Psi^{(+)}_{KJ\lam_J\{s_k\lam_{s_k}\}}=
\left[
\prod_{k=1}^{3}
(1/m)
\bar{u}^{(\lam_{s_k})}_{\al_k}(\p_k)
\right]
\Psi_{KJ\lam_J}^{\{\al_k\}}(\{p_k\})=
(2\pi)
\delta^{(1)}(\stackrel{o}{P}_{0}-M)
\la
\{\p_k s_k \lam_{s_k}\}
\mid K J \lam_J \ra.
\label{psi_polozh_chast}
\ee
The function~(\ref{psi_polozh_chast}) is named a wave function projected
onto positive frequency states~\cite{Faustov1}.
Taking into account~(\ref{fourie_equal_times_WF})
and~(\ref{psi_chrz_compon}) we obtain:
\be
\la
\{\p_k s_k \lam_{s_k}\}
\mid K J \lam_J \ra.
=
(2\pi)^3
\delta^{(3)}(\sum_{k=1}^3\po_k)
\left[
\prod_{k=1}^{3}
\stackrel{+}{D}\smash{^{1/2}_{\lam'_{s_k}\lam_{s_k}}}(L^{-1}_K,p_k)
\right]
\Phi^{(+)}_{MJ\lam_J\{s_k\lam'_{s_k}\}}(\{\po_k\}),
\label{phiplus_chrz_comp}
\ee
where
$\Phi^{(+)}_{MJ\lam_J\{s_k\lam'_{s_k}\}}
$
is related with
$\Phi^{\al_k}_{MJ\lam_J},
$, defined in~(\ref{fourie_equal_times_WF})), via:
\be
\Phi^{(+)}_{MJ\lam_J\{s_k\lam'_{s_k}\}}(\{\po_k\})=
\left[
\prod_{k=1}^{3}
(1/m)
\bar{u}^{(\lam_{s_k})}_{\al_k}(\p_k)
\right]
\Phi^{\al_l}_{MJ\lam_J}(\{\po_k\}).
\ee
Using the method of the two-time Green function~\cite{Log-Tav63,Faustov1}
it is possible to obtain quasipotential equation
for function
$
\Phi^{(+)}
$:
\be
\frac{1}{(2\pi)^6 E_{\po_3}}
(M_0-M)
\Phi^{(+)}_{MJ\lam_J\{s_k\lam_{s_k}\}}(\{\po_k\})=
\int
d\Omega_{\po'_1}d\Omega_{\po'_2}
V_{\{\lam_{s_k}\lam'_{s_k}\}}(M,\{\po_k|\po'_k\})
\Phi^{(+)}_{MJ\lam_J\{s_k\lam'_{s_k}\}}(\{\po'_k\}).
\label{QPE_for_phiplus}
\ee
In the impulse approximation the quasipotential
$V_{\{\lam_{s_k}\lam'_{s_k}\}}(M,\{\po_k|\po'_k\})$
may be independent of the energy,
because of the nucleon mass in the equation
iteration~(\ref{QPE_for_phiplus})
may be replaced in quasipotential by sum of the
free quark energy in c.m.s.
$M_0(\{\po_k\})=\sum_{k=1}^3E_{\po_k}$
and then the next order in the coupling constant
may be neglected.
The wave function normalization condition obtained
by means of the Green function~\cite{Faustov1}
for the quasipotential independent of energy
has a symmetric form relatively the particle permutations:
\be
\frac{1}{(2\pi)^6}
\int
d\Omega_{\po_1}d\Omega_{\po_2}
\mid\Phi^{+}_{MJ\lam_J\{s_k\lam_{s_k}\}}(\{\po_k\})
\mid^2
\frac{1}{ E_{\po_3}}=1.
\label{norm_phiplus}
\ee

\newpage
\begin{center}
{\bf 2. Nucleon spin wave function.}
\end{center}
\par
Now let us use the general method
to construct the state with a definite momentum in the system of
free particles to study spin dependence of QP WF by using
connection of QP WF with a decomposition component of the nucleon
vector state in the Fock-momentum space~(\ref{phiplus_chrz_comp}).
In the center-of mass system formula~(\ref{phiplus_chrz_comp}) has the form:
\be
\la\{\po_k s_k\lam_{s_k}\}| M\Ko J \lam_J\ra=
(2\pi)^3\delta^{(3)}(\sum_{i=1}^3\po_i)
\Phi^{(+)}_{J\lam_J\{s_k\lam_{s_k}\}}(\{\po_k\}).
\label{vydelenie_CMS_phiplus}
\ee
Now we decompose $|\{\po_k s_k\lam_{s_k}\}\ra$
over free particle states with a definite momentum in the c.m.s.
$| M_0(\{\po_k\})\sum_k\po_k J\lam_J(\ldots)\ra$.
The dots mean other observables from the complete set which
will be indicated below. Its choise depends on the ways of summing
spins and orbital angular momenta.
Further we use the method of work~\cite{Shirokov58}
to decompose the direct product
of irreducible representations of the inhomogeneous Lorentz group.
\par
In case of the two free particle system in its c.m.s. with
3-momentum
$\ponot_1$ and $\ponot_2$
($\plo \equiv \ponot_1 + \ponot_2 = 0$) and
with the invariant system [1+2] mass
$M_{12} \equiv E_{\ponot_1}+E_{\ponot_2}$,
this decomposition has the following form:
\ben
 |\ponot_1 s_1\lam_{s_1}\ponot_2 s_2\lam_{s_2} \ra =
| M_{12} \plo j \lam_j(ls)\ra
\la j \lam_j | l \lam_l s \lam_s \ra
  Y_{l \lam_l}(\hat{\pmo})
\la s \lam_s | s_1 \lam_{s_1} s_2 \lam_{s_2} \ra.
\label{Shirokov2}
\een
Here $\hat{\pmo}$ is angle variables of momentum
$\pmo$, $Y_{l \lam_l}$ is spherical harmonics,
$l$ is the relative orbital momentum of the system [1+2],
$s=0,1$ is spin,
$j$ is the total spin of system [1+2].
Going from the rest system [1+2]
to c.m.s. of nucleon, we have from~(\ref{Shirokov2}):
\be
| E_{\po_1}+E_{\po_2} \po_{12} j\lam_j (ls)\ra=
\hat{U}(L_{\po_{12}})| M_{12} \plo\ = \0 j\lam_j (ls)\ra,
\label{cherez_u}
\ee
where
$\poc_{12} = \poc_1 + \poc_2$,
$\poc_k=(L_{\ponotc_{12}}\ponotc_k)$ = $(L_K^{-1}p_k)$.
If we introduce the notation
$
y^{[12]l}_{j \lam_j s \lam_s}(\hat{\pmo}) \equiv
\la j \lam_j | l \lam_l s \lam_s \ra
Y_{l \lam_l}(\hat{\pmo})
$, then applying~(\ref{vig_pov1}) we obtain:
\ben
 |\po_1 s_1\lam_{s_1}\po_2 s_2\lam_{s_2} \ra =
| E_{\po_1}+E_{\po_2} \po_{12} j\lam_j (ls)\ra
 y^{[12]l}_{j \lam_j s \lam_s}(\hat{\pmo})
\times
 \nonumber \\
\times
\la s \lam_s | s_1 \lam'_{s_1} s_2 \lam'_{s_2} \ra
\Dppo(L_{\po_{12}},\pmo)
\Dppd(L_{\po_{12}},\pno).
\label{Shirokov3}
\een
Multiplying~(\ref{Shirokov3}) on the vector state of the third quark,
we have obtained:
\ben
 |\{\po_k s_k\lam_{s_k}\} \ra =
| E_{\po_1}+E_{\po_2} \po_{12} j\lam_j (ls) \po_3 s_3\lam_{s_3}\ra
 y^{[12]l}_{j \lam_j s \lam_s}(\hat{\pmo})
\times
 \nonumber \\
\times
\la s \lam_s | s_1 \lam'_{s_1} s_2 \lam'_{s_2} \ra.
\Dppo(L_{\po_{12}},\pmo)
\Dppd(L_{\po_{12}},\pno).
\label{Shirokov4}
\een
In analogy with~(\ref{Shirokov2}) we obtain
(notation $ y^{[[12]3]L}_{J \lam_J S \lam_S}(\hat{\po_3}) =
 \la J \lam_J | L \lam_L S \lam_S \ra
 Y_{L \lam_L}(\hat{\po_3})$ is introduced):
\ben
| E_{\po_1}+E_{\po_2} \po_{12} j\lam_j (ls);\po_3 s_3\lam_{s_3}\ra =
| M_0(\{\po_k\}) \sum_k\po_k J \lam_J(ls,LS)\ra
 y^{[[12]3]L}_{J \lam_J S \lam_S}(\hat{\po_3})
\la S \lam_S | j \lam_j s_3 \lam_{s_3} \ra,
\label{Shirokov5}
\een
where $S$ is the spin of the system [[1+2]+3],
obtained by summation of $j$ and $s_3$;
$L$ is the orbital angular momentum of the
third quark with respect to the [1+2] system,
$J$ is the total spin of the system [[1+2]+3],
obtained by summmation of $S$ and $L$.
Thus, the desired decomposition has the form:
\ben
 &&|\{\po_k s_k\lam_k\} \ra =
 | M_0(\{\po_k\})\sum_k\po_k J\lam_J(ls,LS)\ra\,\,\,
 y^{[[12]3]L}_{J \lam_J S \lam_S}(\hat{\po_3})
 y^{[12]l}_{j \lam_j s \lam_s}(\hat{\pmo})
\times
 \nonumber \\
&&
\times
\la S \lam_S (j s_3)| j \lam_j (ls) s_3 \lam_{s_3} \ra
\la s \lam_s | s_1 \lam'_{s_1} s_2 \lam'_{s_2} \ra
\Dppo(L_{\po_{12}},\pmo)
\Dppd(L_{\po_{12}},\pno).
\label{Shirokov6}
\een
We have separated motion of c.m.s. in the scalar product of the bound system
vector state and the free particle vector state:
\ben
 \la M \Ko = \0 J \lam_J|M_0(\{\po_k\})\sum_k\po_k J\lam_J(ls,LS)\ra =
 (2\pi)^3\delta(\sum_k\po_k) A_{M J}^{*(ls,LS)}(M_0(\{\po_k\})).
\label{ScalProd_deltaFunc}
\een
Independence of $A$ on $\lam_J$ is the result the symmetry
relatively the reflection.
The relation~(\ref{ScalProd_deltaFunc}) was obtained by using unit operator
$\hat{1}=(1/V)\int_V d^3x
\exp^{-\hat{\bf P}{\bf x}}
\exp^{ \hat{\bf P}{\bf x}}$.
According to~(\ref{vydelenie_CMS_phiplus}),~(\ref{Shirokov6})
 and~(\ref{ScalProd_deltaFunc}), for relative motion WF
$\Phi^{(+)}$ in full record with isospin indices we have obtained:
\ben
\Phi^{(+)*}_{MJ\lam_J T\lam_T \{s_k\lam_{s_k}t_k\lam_{t_k}\}}
(\{\po_k\}) =
 A_{M J T\lam_T\{t_k\lam_{t_k}\} }^{*(ls,LS)}(M_0(\{\po_k\}))
y^{[[12]3]L}_{J \lam_J S m_S}(\hat{\po_3})
y^{[12]l}_{j \lam_j s m_s}(\hat{\pmo})
\times
 \nonumber \\
\times\la S \lam_S | j \lam_j s_3 \lam_{s_3} \ra
\la s \lam_s | s_1 \lam'_{s_1} s_2 \lam'_{s_2} \ra
\Dppo(L_{\po_{12}},\pmo)
\Dppd(L_{\po_{12}},\pno).
\label{Shirokov7}
\een
Since the contributions of higher orbital moments in nucleon
are small~\cite{SimulaPR2001},
we write only the S-wave part ($L=l=0$)
omitting spin, isospin and symbols $L=l=0$:
\ben
\Phi^{(+)}_{M
    \{\lam_{s_k}\lam_{t_k}\}
          } (\{\po_k\}) =
 \DDo(L_{\po_{12}},\pmo)
 \DDd(L_{\po_{12}},\pno)
 \delta_{\lam_{s_3}\lam'_{s_3}}
 \xi^{j}_{\{\lam'_{s_k}\}}
 A_{M\{\lam_{t_k}\}}^{(j)}(M_0(\{\po_k\})),
 \label{Shirokov9}
\een
where $\xi^{j}_{\{\lam_{s_k}\}} =
\sum_{j=0,1} \la \lam_{s_1}\lam_{s_2} |j \lam_j \ra
\la j \lam_j\lam_{s_3} |\lam_J
\ra $.
Let us decompose functions $A_{M\{\lam_{t_k}\}}^{(j)}$
over two isospin basic functions $\eta^{\tau}_{\{\lam_{t_k}\}}$;
$\tau=0,1$ is the [1+2] system isospin :
$A_{M\{\lam_{t_k}\}}^{(j)}=B_M^{(j,\tau)}\eta^\tau_{\{\lam_{t_k}\}}$
(sum over $\tau$) and substitute it in~(\ref{Shirokov9}):
\ben
\Phi^{(+)}_{\{\lam_{s_k}\lam_{t_k}\} }
(\{\po_k\})  =
\left[
\DDo(L_{\po_{12}},\pmo)
\DDd(L_{\po_{12}},\pno)
\delta_{\lam_{s_3}\lam'_{s_3}}
\right]
\xi^{j}_{\{\lam'_{s_k}\}}
 B_{M}^{(j,\tau)}(M_0(\{\po_k\}))
\eta^{\tau}_{\{\lam_{t_k}\}}.
\label{phiplus_symm}
\een
\par
The function $\Phi^{(+)}$~(\ref{Shirokov7})
is symmetrical relatively the particle permutations
and with antisymmetric colour fuction realizes
antisymmetrical representation of the permutations group.
S-wave function~(\ref{phiplus_symm})
with zero-orbital momentum of the [1+2] system
and zero angular orbital momentum of the third quark with respect
to the [1+2] system, is antisymmetric relatively the quark permutations.
Let us suppose, that interquark interaction does not depend on
the spin and isospin:
$B_M^{(j,\tau)}(M_0(\{\po_k\})) =
(2\pi)^3C^{(j,\tau)}\varphi_M(M_0(\{\po_k\}))/\sqrt{2}$ and
we choose the normalization constants to agree with the 
nonrelativistic theory:
$C^{(0,0)}=C^{(1,1)}=1$, $C^{(1,0)}=C^{(0,1)}=0$.
Introducing the notation for the symmetric spin-isospin wave function
$
\chi_{\{\lam_{s_k}\lam_{t_k}\}}=
\left[
\xi^{1}_{\{\lam_{s_k}\}}
\eta^{1}_{\{\lam_{t_k}\}}
+
\xi^{0}_{\{\lam_{s_k}\}}
\eta^{0}_{\{\lam_{t_k}\}}
\right]/\sqrt{2}
$
, we obtain
\ben
\Phi^{(+)}_{\{\lam_{s_k}\lam_{t_k}\} }
(\{\po_k\})  =
(2\pi)^3
\left[
\DDo(L_{\po_{12}},\pmo)
\DDd(L_{\po_{12}},\pno)
\delta_{\lam_{s_3}\lam'_{s_3}}
\right]
\chi_{\{\lam'_{s_k}\lam_{t_k}\}}
\varphi_M(M_0(\{\po_k\}))
\label{Shirokov91}
\een
\par
Two $D$-matrices in~(\ref{Shirokov91})
represent the minimal kinematic violation of the $SU(6)$
symmetry. The general expression for WF~(\ref{Shirokov7})
indicates three ways of
$SU(6)$ violations, related to relativization of the model:
(1) including the admixture of the mixed $SU(6)$ symmetry in WF,
(2) taking into account the quark interaction dependence
on spin and isospin,
(3) and  P- and D-waves.

\begin{center}
{\bf 3. Effective mass approximation and relativistic oscillator.}
\end{center}
\par
According to~(\ref{decomp_1}),~(\ref{norm_phiplus})
 and~(\ref{Shirokov91}),
function $\varphi_M$ satisfies the equation
and normalization condition:
$$
[M_0(\{\po_k\})-M]
\varphi_M(M_0(\{\po_k\}))=
\int
d\Omega_{\po'_1}d\Omega_{\po'_2}
v(\{\po_k|\po'_k\})
\varphi_M(M_0(\{\po_k\}))/E_{\po'_3},
$$
\ben
\int
d\Omega_{\po_1}d\Omega_{\po_2}
\mid
\varphi_M(M_0(\{\po_k\}))
\mid^2
/E_{\po_3}=1,
\label{QPE_varphi}
\een
where $v$ is a quasipotential determined by relation:
$v=(2\pi)^6[\chi(DDI)V(DDI)\chi]E_{\po'_3}E_{\po_3}$,
where $(DDI)$ is a matrix record of factors
in square brackets of expression~(\ref{Shirokov91}).
\par
Variables $\po_1$, $\po_2$ and $\po_3$ are equivalent and 
the choise  of variables $\po_1$, $\po_2$ as independent 
does not change the three quark
equivalent to calculate the average values.
For example, the three effective quark masses, obtained by formula
$m^{eff}=\sqrt{m^2+\la p^2_k\ra}$ are equal
($\la p^2_k\ra=\int
d\Omega_{\p_1}d\Omega_{\p_2}
\mid
\varphi_M
\mid^2
p^2_k
/E_{\po_3}
$).
The approximate relativistic models, using the idea of the 
effective quark mass,
were considered in work~\cite{Lucha,Gerasimov82}.
In these works the effective quark mass
is introduced as a parameter instead of the quark mass. 
In our work the effective
quark mass has been introduced as a suitable approximation.
\par
Let us pass to the semi-momenta~\cite{Skachkov_Solovcov78}
in~(\ref{QPE_varphi}):
$
\pi_k=\po_k
\sqrt{2m/(m+E_{\po_k})},
\;
E_{\po_k}=m+\pi_k^2/2m.
$
Equation~(\ref{QPE_varphi}) takes a nonrelativistic form in terms of $\pi_k$.
In equation~(\ref{QPE_varphi}) we rewrite the quasipotential as an analog of nonrelativistic oscillator:
\be
\left[
\pi_1^2/2m+\pi_2^2/2m+\pi_3^2/2m+3m-M-
k_o(\nabla^2_{\pi_1}+\nabla^2_{\pi_2}-\nabla_{\pi_1}\nabla_{\pi_2})
\right]
\varphi_M^{osc}=0.
\label{QPE_osc_pi}
\ee
Using new variables
$
\bk=\frac{1}{2}(\pi_1-\pi_2),
\;
\bk'=\frac{1}{3}(\pi_1+\pi_2-2\pi_3),
\;
\Q=\pi_1+\pi_2+\pi_3
$
and  parameters in~(\ref{QPE_osc_pi})
$\mu=m/2,\;\mu'=2m/3,\;\omega^2=3k_o/m$, we have obtained:
\be
\left[
3m-M+
\bk^2/2\mu+\bk'^2/2\mu'+\Q^2/6m
-
(\mu/2)\omega^2\nabla_{\bk}^2
-
(\mu'/2)\omega^2\nabla_{\bk'}^2
\right]
\varphi_M^{osc}=0.
\label{QPE_osc_k_kstr}
\ee
Decomposing the quark energies in semi-momenta in the $m^{eff}$
neighborhood
and taking into account the condition $\sum_{k=1}^3\po_k=0$, 
it is possible to
show that term $\Q^2/6m$ has a small contribution into the energy, 
if value
$\delta_E\equiv m^{eff}-\la E_{p_k}\ra$ is small
(the effective mass aproximation). 
In the zero-order approximation relatively the
$\Q^2/6m$, we obtain a solution of equation~(\ref{QPE_osc_k_kstr})
for the ground state
($\lam^2=\gamma^2/2,
\;
\lam'^2=2/3\gamma^2,
\;
\gamma^2=m\omega
$):
$
\varphi_M^{osc}\approx\exp(-k^2/2\lam^2-k'^2/2\lam'^2).
$
Taking into account the effective mass approximation,
this solution is reduced to the form:
\be
\varphi_M^{osc}\approx N\exp[-m(M_0-3m)/\gamma^2].
\label{varphi_approx_eff_m}
\ee
\par
Numerical calculations with the relativistic oscillator
WF~(\ref{varphi_approx_eff_m}) gives
$m^{eff}/m\in[1.08,1.74]$ for $\gamma/m\in[0.4,1.1]$.
In this case $\delta_E$ is positive and is in limits
$[0.002,0.009]m^{eff}$.
Smallness of $\delta_E$ means the smallness of the momentum module dispersion.
Indeed, in the ground state
$\la\p_k^2-\la\p_k^2\ra\ra\ra=0$ and therefore
$\delta_E\approx \sigma^2_{\p_k^2}/8(m^{eff})^3$,
where $\sigma^2_{\p_k^2}=\la[\p_k^2-\la\p_k^2\ra]^2\ra.$
\par
We note that, WF in form $\exp[-M^2_0/\gamma^2]$, applied in many works,
is different from~(\ref{varphi_approx_eff_m})
by the order of $M_0$ and has different asymptotic
behaviour for high momenta.

\begin{center}
{\bf Conclusion.}
\end{center}
\par
In the impulse approximation, represented by the zero-order field operators
in the Bethe-Salpeter WF, the quasipotential wave function reduces
to the three-quark component of the vector state decomposition
of the bound system in the Fock-momentum space.
This allows  us to apply the standart method for the decomposition 
of the irreducible representations product
of the inhomogeneous Lorentz-group
over states with the definite momentum to analyse of the QP WF structure.
The physical preference of the $SU(6)$-symmetric solution is determined by
its uniqueness in the nonreletivistic limit.
\par
The model of the relativistic three-particle oscillator being the
direct generalization
of the nonrelativistic oscillator is proposed.
Numerically it was shown that the effective mass approximation may be applyed
in the wide region of the oscillator parameters,
in which the ratio of the momentum dispersion to the average value 
of the quark momentum module is small.
\par
Authors express gratitude to N.V.Maksimenko for support and E.A.Dey
for usefull remarks.

\end{document}